\documentclass[aps,prb,showpacs,reprint,superscriptaddress]{revtex4-2}

\usepackage{graphicx}
\usepackage{color}
\usepackage{amsmath}
\usepackage{bbm}
\usepackage{amssymb}
\usepackage{nameref}
\usepackage{hyperref}
\usepackage{braket}
\usepackage{xcolor}

\usepackage{dsfont}

\usepackage[utf8]{inputenc}	
\usepackage[T1]{fontenc}
\usepackage{bm}

\usepackage[svgnames]{xcolor}

\input epsf

\begin{document}


\title{Emergent Quantum Valley Hall Insulator from Electron Interactions in Transition-Metal Dichalcogenide Heterobilayers}

\texorpdfstring{}


\author{Palash Saha}
\affiliation{Academic Centre for Materials and Nanotechnology, AGH University of Krakow, Al. Mickiewicza 30, 30-059 Krakow, Poland}
\author{Micha{\l} Zegrodnik}%
 \email{michal.zegrodnik@agh.edu.pl}
\affiliation{Academic Centre for Materials and Nanotechnology, AGH University of Krakow, Al. Mickiewicza 30, 30-059 Krakow, Poland}%

\date{\today}

\begin{abstract}
We explore the emergence of topological phases in moir\'{e} MoTe$_2$/WSe$_2$ bilayer, highlighting the crucial role of spin-orbit coupling and Coulomb interactions at two holes per moir\'e unit cell \(v = 2\). Our analysis uncovers robust Quantum Valley Hall Insulating (QVHI) phase and reveals that long-range interactions alone can mediate the interlayer electron tunneling, generating topologically nontrivial bands even in the absence of the corresponding single-particle hopping. Additionally, we show that in the case of band mixing terms originating both from the interaction and single particle physics a competition between topological states realizing $s$-$wave$ and $p\pm ip$-$wave$ symmetries can appear. Moreover, within the considered theoretical framework, we present that by introducing a small Zeeman field, one can lift the band inversion in one of the valleys. This leads to a Quantum Anomalous Hall Insulating (QAHI) state with the topological gap opening in a single valley and the other being topologically trivial. 

\end{abstract}

\maketitle

\section{\label{sec:level1}Introduction}

The exploration of topologically nontrivial and exotic states in moir\'{e} superlattices has recently become a key objective in condensed matter physics. In particular, transition metal dichalcogenide (TMD) bilayers due to lattice mismatch or rotational misalignment have lead to experimental observation of 
topologically protected edge states \cite{exp_top_1,exp_top_2,exp_top_3}, 
unconventional superconductivity
\cite{exp_sc_1,exp_sc_2}, 
correlated insulating states
\cite{exp_insul_1,exp_insul_2,exp_insul_3}
as well as 
generalized Wigner crystals
\cite{exp_wig_1,exp_wig_2}.
Due to the presence of narrow electronic bands, it is believed that effects related with electron-electron interactions may play a significant role in the emergence of the variety of observed physical phenomena.

One of the characteristic features of the TMD-based bilayer systems is the presence of strong Ising-type spin-orbit coupling, which leads to spin-valley locking\;\cite{sv_lock_1,sv_lock_2,sv_lock_3}. In the AB stacked MoTe$_2$/WSe$_2$ heterobilayer, a nonzero band offset separates the spin-valley locked bands originating from the upper and lower layer, leading to a band insulator at two holes per moir\'{e} unit cell \(v = 2\). By applying a perpendicular electric field, one can significantly reduce the band gap, enabling the band inversion necessary for a topological phase transition\;\cite{louk,band_inv_2}. Such physical picture is consistent with the experimentally observed Quantum Valley Hall Insulator (QVHI)\;\cite{exp_1,b_field}. Furthermore, at one hole per moir\'{e} unit cell \(v = 1\) (half-filling of the upper band), a Mott (or charge transfer) insulator emerges, which is due to the strong Coulomb repulsion. In this case, after the application of the displacement field a transition to a Quantum Anomalous Hall Insulator (QAHI) is observed\;\cite{exp_1}.

First-principles calculations and the continuum model approach have been utilized to search for QVHI phases in MoTe$_2$/WSe$_2$ heterobilayer\;\cite{louk,chern_4}. Also, interaction driven quantum valley Hall effects have been studied in the continuum limit using plane wave basis \cite{plane_wave} and pseudomagnetic fields \cite{pseudo_magnetic}. The role of electron-electron interactions has also been emphasized in the analysis of topological bands of the QAHI state \cite{band_inv_2,insu_2,pseudo_magnetic, chern_1,chern_2,chern_4} as well as heavy-fermion/Kondo \cite{kondo_1,kondo_2}
and excitonic \cite{exciton_1,exciton_2}
physics. These studies suggest that single-particle physics alone is insufficient to capture the low-energy behavior of the MoTe$_2$/WSe$_2$ system.

Our work is motivated by the experimental findings \cite{exp_1,b_field} that demonstrate a continuous topological phase transition from a  moir\'{e} band insulator to a QVHI at two holes per moir\'{e} unit cell in the AB-stacked MoTe$_2$/WSe$_2$ heterobilayer. It should be noted that in the lowest order continuum model of this system, the interlayer spin-flip hopping which is necessary to induce the topological features is vanishingly small\cite{chern_4,Qimao_2024,exp_top_3}. Here, we apply a minimal model composed of two moir\'{e} valence bands, supplemented by intra- and inter-site Coulomb repulsion terms treated with the use of Hartree-Fock method (HF). We show that Coulomb interaction-assisted interlayer hopping can effectively replicate the crucial spin-flip tunneling, leading to a non-trivial band topology consistent with QVHI state. We calculate the winding number of this topological gap's complex phase in the neighborhood of nodal points which determine the topological features of the model. Moreover, in the presence of both interaction induced band mixing and small single particle mixing term a competition between two topological states realizing different symmetries can appear. Additionally, we show that within the presented theoretical framework one can lift the spin-valley degeneracy in the $\mathcal{T}$ invariant QVHI \cite{km1} state by using the Zeeman field as discussed in the recent experimental report \cite{b_field}. In such a case, the band inversion and emergence of topological bands become confined to one valley, leaving the other topologically trivial. Ultimately, this presents a route to realize \(\mathcal{T}\) symmetry broken QAHI from QVHI at \(v\) = 2 via the interplay between the external magnetic field and electron-electron interaction effects. Note that although an external magnetic field is applied, it does not result in a conventional Quantum Hall effect due to absence of Landau level formation \cite{haldane}.

The paper is structured as follows:  In Sec. II, we introduce the extended Hubbard model and outline the implementation of the HF approach. Sec. III explores in detail the topological properties of the considered system as induced by the interlayer interaction term.
In the second part of Sec. II, we study the influence of Zeeman field in bridging the QVHI and QAHI phase. Finally, in Section. IV we summarize and conclude the important results. 


\section{Model And Method} 

As shown in Ref. \cite{louk}, the minimal model of the AB-stacked MoTe$_2$/WSe$_2$ can be composed of two moir\'{e} Wannier orbitals centered at the $XX$ ($X=$Te, Se) and $MM$ ($M=$Mo, W) stacking points in the WSe$_2$ and MoTe$_2$ layers, respectively. In order to take into account electron-electron interactions we supplement the model with both onsite and intersite Coulomb repulsion terms, which results in the extended Kane-Mele-Hubbard Hamiltonian of the form
\begin{equation}
\hat{\mathcal{H}} = \hat{\mathcal{H}_t}+\hat{\mathcal{H}}_{UV}+ \hat{\mathcal{H}}_{B},
\label{ham}
\end{equation}
where the first term represents the single-particle part, shown explicitly below
\begin{equation}
\begin{split}
\hat{\mathcal{H}}_{t} &= 
\sum_{\langle\langle ij\rangle\rangle l\sigma} t^{ll}_{ij\sigma}\;\hat{c}_{i l \sigma}^{\dagger}\; \hat{c}_{j l \sigma}^{}
+ \sum_{\langle ij\rangle \sigma} \big(t^{12}_{ij\sigma\bar{\sigma}}\;\hat{c}_{i 1 \sigma}^{\dagger}\; \hat{c}_{j 2 \bar{\sigma}}^{}+H.c.\big)\\
&+(D+\Delta)\sum_{i}\hat{n}_{il=2},
\end{split}
\label{ham_t}
\end{equation}
with $\hat{c}_{i l \sigma}^{\dagger}$ ($\hat{c}_{i l \sigma}$) being the creation (annihilation) operator for an electron of spin $\sigma$ in the Wannier orbital at site $i$ where $\bar{\sigma}$ denotes the spin opposite to $\sigma$ and $\langle\langle ij\rangle\rangle$ denotes next-nearest-neighbor (NNN) pairs on the honeycomb lattice (intralayer hoppings) while $\langle ij\rangle$ denotes  nearest-neighbor (NN) pairs (interlayer hoppings). Above, we introduced a layer index \(l\) to distinguish between \(MM\) (\(l=1\)) and \(XX\) (\(l=2\)) orbitals corresponding to two triangular sublattices, which together form a honeycomb structure. It should be noted that high values of spin-orbit interaction in TMDs lead to a strong coupling between spin and valley degrees of freedom, effectively locking them together. When MoTe$_2$ and WSe$_2$ are stacked in an AB configuration, the spin-valley locking exhibits an opposite orientation in the two layers. In the minimal picture considered here, this leads to complex hoppings of the following form 
\begin{equation}
    t^{ll}_{ij\sigma}=t_l\; e^{i \phi_{\parallel} \sigma^z \nu_{ij}} 
    \label{hopping_amp_intralayer}
\end{equation}
\begin{equation}
    t^{12}_{ij\sigma\bar{\sigma}}=t_{\perp}\; e^{i \phi_{\perp} \sigma^z \eta_{ij}},
    \label{hopping_amp_interlayer}
\end{equation}
where, $\sigma$ = \{$\uparrow,\downarrow$\} labels the electron spin index, while $\sigma^z=1$ $(\sigma^z=-1)$ corresponds to $\sigma=\uparrow (\sigma=\downarrow)$.
Interorbital amplitude in Eq. (\ref{hopping_amp_interlayer}) describes a spin-flip process between two different orbitals of our model where the value of $\sigma_z$ is defined based on the final state $\sigma$ of the electron.
\begin{figure}[t]
\centering
\includegraphics[width=0.9\linewidth, height=3.2cm]{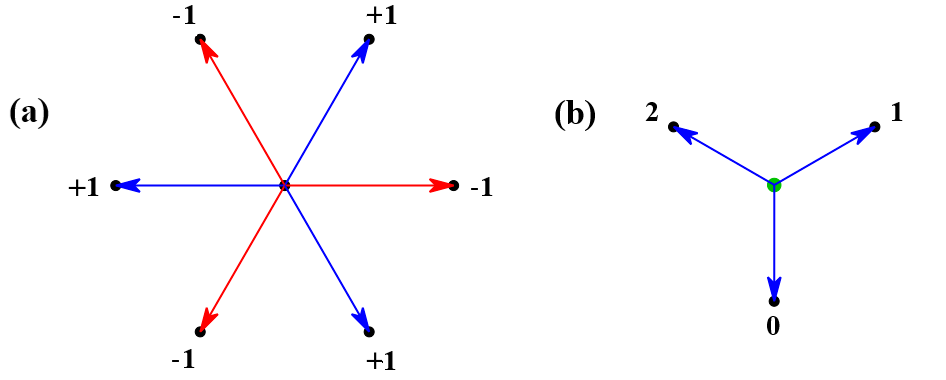}
\caption{(a) The direction dependent sign parameter $\nu_{ij}=\pm 1$ for the six NNN hoppings appearing in Eq. (\ref{hopping_amp_intralayer}) , (b) The  $\eta_{ij}=0,1,2$ parameters for the three NN hoppings appearing in Eq. \ref{hopping_amp_interlayer}.}
\label{arrows}
\end{figure}
The parameter $\nu_{ij} = \pm1$ is determined by the intralayer hopping direction (cf. Fig. \ref{arrows}) while $\eta_{ij} = 0,1,2$ enumerates different interlayer hopping directions, increasing counterclockwise around the WSe$_2$ lattice sites. Finally, the complex phases of the hoppings are $\phi_{\parallel}=\phi_{\perp}=2\pi/3$. However, for the sake of completeness, in the following we also consider the case of $\phi_{\parallel}=2\pi/3$ and $\phi_{\perp}=0$. The structure of the single-particle part of the Hamiltonian is analogical to that of the Kane-Mele model, however, to obtain the latter from the former one has to carry out a unitary transformation which flips the spin on one of the sublattices. Also, in the Kane-Mele model one typically takes $\phi_{\parallel}=\pi/2$ and $\phi_{\perp}=0$. The hopping amplitudes in our approach are adopted from Ref.~\onlinecite{louk} and have the values $t_1 = 4.03$ meV and $t_2 = 3.4$ meV. 
As discussed in Appendix A, the rise of spin-flip hybridization leads to an immediate effect contributing to the loss of flavor symmetry. In the following section, we focus on the $t_{\perp}=0$ case; however, for the sake of generality, the influence $t_{\perp}\neq 0$ is also discussed.

The third term in Eq. (\ref{ham_t}) represents the onsite contribution, where the particle number operator is defined as \(\hat{n}_{il} = \hat{n}_{il\uparrow} + \hat{n}_{il\downarrow}\) with \(\hat{n}_{il\sigma} = \hat{c}_{il\sigma}^{\dagger} \hat{c}_{il\sigma}\). Here, \(\Delta = -60\) meV characterizes an intrinsic band offset, while \(D\) accounts for the influence of the displacement field. When \(D = 0\), the top and bottom bands are primarily formed by MoTe\(_2\) and WSe\(_2\) states, respectively, with a band gap between them. As \(D\) increases, the bottom band shifts upward, eventually leading to band inversion\cite{louk}.

The second term of Hamiltonian (\ref{ham}) corresponds to electron-electron interactions and has the form
\begin{equation}
\hat{\mathcal{H}}_{UV} = 
U\sum_{il}{ \hat{n}_{il\uparrow} \hat{n}_{il\downarrow} }
+ V_{} \sideset{}{^{\prime}}\sum_{\substack{\langle ij ll'\rangle\sigma}}{ \hat{n}_{il} \hat{n}_{jl'}},
\label{ham_UV}
\end{equation}
where $U$ and $V_{}$ correspond to onsite and intersite Coulomb repulsion integrals and the primed summation means each NN neighbor bond is taken in account only once. The precise realistic value of $U$ is difficult to determine exactly and can differ from one experimental setup to the other due to the influence of dielectric environment of the sample. Also, by applying a twist angle between the two monoatomic layers one can tune both the band width ($W$) and the value of $U$. The observation of Mott (or charge transfer) insulator for one hole per moir\'e unit cell of the AB-stacked, angle aligned system\;\cite{exp_1} indicates the strongly correlated regime, meaning that the value of $U$ should be at least larger than the width of the upper band, $U>W$ (in our case $W\approx25-35\;$meV depending on the displacement field). On the other hand, the signatures of the Wigner-like insulating states have not been observed at fractional fillings of this particular moir\'e structure. This would mean that the $V$-term is not extremely strong ($V\lesssim W$)\cite{bibotski_2025}. The standard Gaussian Wannier approximation can give us a rough estimate of the onsite Coulomb repulsion in the form $U\approx e^2\sqrt{\pi}/(4\pi\epsilon_0\epsilon_{\mathrm{eff}}l)$. For the angle aligned system with orbital width $l\approx2\;$nm (the lattice constant $a_M=5.5\;$nm) and $\epsilon_{\mathrm{eff}}\approx 8-12\;$ (for hBN encapsulated system with nearby metallic gates) we obtain $U\approx100-160\;$meV. Within the same approximation we get $U/V\approx\sqrt{\pi}a_M/l$, which leads to  $V\approx20-30\;$meV. In our considerations, if not stated otherwise, we focus on the situation with $U= 94.7\;$meV and analyze the properties of the system as a function of the intersite Coulomb coupling, $V$, which plays a crucial role in the current work. Other theoretical analysis consider, e.g., $U=70\;$meV and $V=U/2$\;\cite{Qimao_2024}. But also much larger strengths of onsite repulsion have been considered with $U=50t$ (for $t=t_1=t_2$)\cite{band_inv_2}, which in our case would roughly correspond to $U=200\;$meV. A large $U$ limit has also been considered in Ref. \cite{exciton_2} where the $t$-$J$ model has been applied supplemented with an intersite repulsion term with $V\lesssim3t$, which would lead to $V\lesssim12\;$meV in our case.

As a part of our analysis we aim at studying the influence of external magnetic field on the topological properties of the system. Therefore, we supplement our model with a Zeeman term of the form
 \begin{eqnarray}
 \hat{\mathcal{H}}_{B}= \mu_B g \sum_\textbf{k} \textbf{B}\cdot\hat{\textbf{S}}(\textbf{k}),
\label{ham_B}
\end{eqnarray}
where \(\mu_B\) is the Bohr magneton, \(g\) is the Land\'{e} g-Factor and 
 \begin{eqnarray}
 \hat{\textbf{S}}(\textbf{k}) = \frac{1}{2}\sum_{\textbf{k}\alpha\beta} \hat{c}^{\dagger}_{\textbf{k}\alpha}\sigma_{\alpha\beta}\hat{c}^{}_{\textbf{k}\beta},
\label{S}
\end{eqnarray}
is the spin operator with \(\sigma\) being the Pauli matrices and \(\alpha,\beta\) = \([\uparrow,\downarrow]\). We focus on the typical situation of out of plane \(\textbf{B}\) with zero inplane magnetic field terms. 

\begin{figure*}[t!]
\centering
\includegraphics[width=0.95\linewidth, height=5cm]{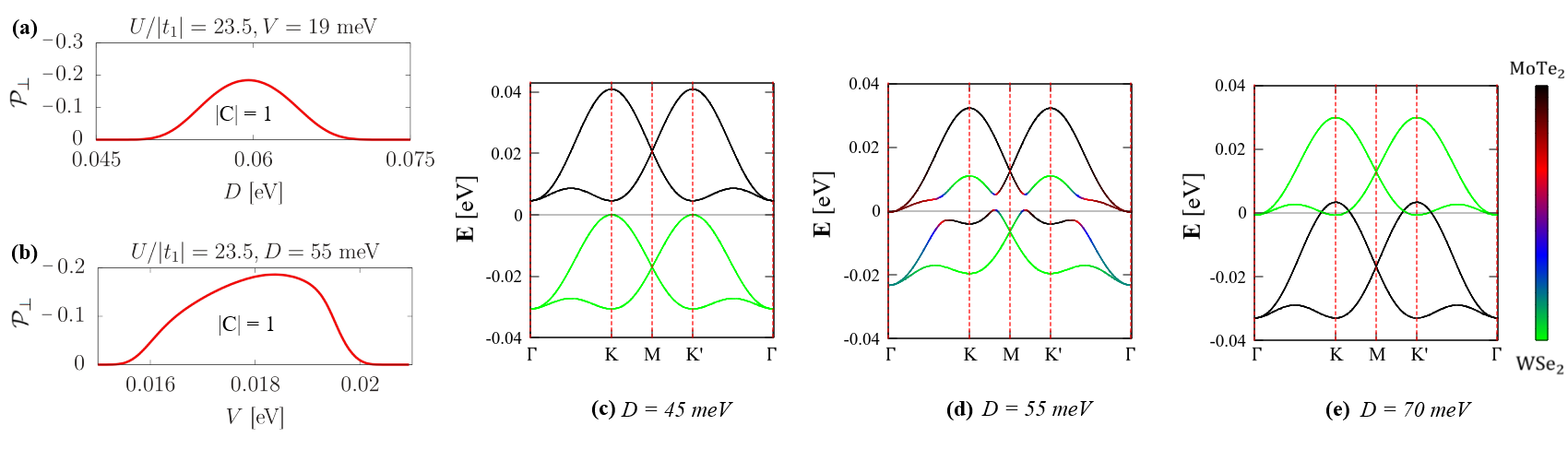}
\caption{(a),(b) The interlayer tunneling expectation value $\mathcal{P}_{\perp}$ as a function of the displacement field and Coulomb interaction for the special case of \(t_{\perp}=0\).
 The band structures in (c), (d), and (e) correspond to the band insulator phase, the QVHI phase, and to the metallic behavior and have been obtained for displacement field values \( D = 45 \)~meV, \( D = 55 \)~meV, and \( D = 70 \)~meV, respectively with the interaction strengths fixed at \( V = 19 \)~meV.
The results have been obtained for $U$ = 94.7 meV and $n=2$, which is equivalent to two holes per moir\'{e} unit cell.}
\label{3_bands}
\end{figure*}
After applying the Hartree-Fock (HF) approximation to the interaction part of the model, 
we obtain
\begin{equation} \begin{split} \hat{\mathcal{H}}_{UV} &\approx U \sum_{il} \left( \langle{\hat{n}}_{il \uparrow}\rangle \hat{n}_{il \downarrow} + \hat{n}_{il \uparrow}\langle \hat{n}_{il \downarrow}\rangle - \langle \hat{n}_{il \uparrow} \hat{n}_{il \downarrow} \rangle\right) 
\quad\\ &+\frac{1}{2}{V} \sideset{}{^{}}\sum_{\substack{\langle ij ll'\rangle\sigma}} \Bigl(\langle{\hat{n}}_{il\sigma}\rangle \hat{n}_{jl'\sigma'} + \hat{n}_{il\sigma} \langle{\hat{n}}_{jl'\sigma'}\rangle - \langle{\hat{n}}_{il\sigma} {\hat{n}}_{jl'\sigma'}\rangle 
\quad\\ &-({{\mathcal{P}}^{ll'}_{i\sigma j\bar{\sigma}}})^{*} \hat{{\mathcal{P}}}_{i\sigma j\bar{\sigma}}^{ll'} - {\mathcal{P}}_{i\sigma j\bar{\sigma}}^{ll'} \hat{{\mathcal{P}}}_{j\bar{\sigma} i{\sigma}}^{l'l} + {\mathcal{P}}_{i\sigma j\bar{\sigma}}^{ll'} {({\mathcal{P}}_{i\sigma j\bar{\sigma}}^{ll'})}^{*}\Bigr) ,
\end{split} 
\label{ham2}
\end{equation}
where $\hat{{\mathcal{P}}}_{i\sigma j\bar{\sigma}}^{ll'} = \hat{c}_{i l \sigma}^{\dagger} \hat{c}_{j l' \bar{\sigma}}^{}$ and ${\mathcal{P}}_{i\sigma j\bar{\sigma}}^{ll'}=\langle{\hat{c}_{i l \sigma}^{\dag} \hat{c}_{j l' \bar{\sigma}}^{}} \rangle$.
\vspace{0.33cm}

In the next step, we rewrite our Hamiltonian in a compact form after the transformation to $\mathbf{k}$-space. Namely,
\begin{equation}
\begin{split}
    \hat{H} &= \sum_{\mathbf{k}\sigma}\hat{\mathbf{f}}^{\dagger}_{\mathbf{k}\sigma}\mathbf{H}_{\mathbf{k}\sigma}\hat{\mathbf{f}}_{\mathbf{k}\sigma}-U  \sum_{il}\langle{\hat{n}_{il\uparrow} \hat{n}_{il\downarrow}}\rangle\\
    &-\frac{V}{2}\bigg( \sum_{\langle ij\rangle\sigma\sigma'}\langle{\hat{n}}_{i1\sigma} {\hat{n}}_{j2\sigma'}\rangle-\sum_{\langle ij\rangle\sigma} 
\mathcal{P}_{i\sigma j\bar{\sigma}}^{12}   \mathcal{P}_{j\bar\sigma i\sigma}^{21} \bigg), 
\end{split}
\end{equation}
where we kept the trivial non-operator terms in the real space representation while $\hat{\mathbf{f}}^{\dagger}_{\mathbf{k}\sigma}=(\hat{c}_{\mathbf{k}1\sigma}^\dagger,\;\hat{c}_{\mathbf{k}2\bar{\sigma}}^\dagger)$ and $\hat{\mathbf{f}}_{\mathbf{k}\sigma}=(\hat{\mathbf{f}}_{\mathbf{k}\sigma}^{\dagger})^{\dagger}$ are composite creation and annihilation operators in $\mathbf{k}$-space and
\begin{equation}
\mathbf{H}_{\mathbf{k}\sigma} =
\begin{pmatrix}
\varepsilon_{\mathbf{k}\sigma}^{11} & \varepsilon_{\mathbf{k}\sigma\bar\sigma}^{12} \\
\varepsilon_{\mathbf{k}\bar\sigma\sigma}^{21} & \varepsilon_{\mathbf{k}\bar\sigma}^{22}
\end{pmatrix}.
\label{eq:matrix_hamiltonian}
\end{equation}
The diagonal elements of the Hamiltonian matrix include single particle contribution ($\sim t^{ll}_{ij}$) and the intraorbital density (Hartree) terms which describe orbital splitting originating from the Coulomb interactions, given by
\begin{equation}
    \varepsilon_{\mathbf{k}{\sigma}}^{11}=\sum_{\langle\langle i(j) \rangle\rangle}t^{11}_{ij\sigma}\;e^{i\mathbf{k}(\mathbf{R}_{i1}-\mathbf{R}_{j1})}+U\langle{\hat{n}_{k1\bar\sigma}}\rangle+3V\langle{\hat{n}_{k2}}\rangle,
    \label{eq:epsilon_11}
\end{equation}
\vspace{-2.5pt} 
\begin{equation}
    \varepsilon_{\mathbf{k}\bar{\sigma}}^{22}=\sum_{\langle\langle i(j) \rangle\rangle}t^{22}_{ij\bar\sigma}\;e^{i\mathbf{k}(\mathbf{R}_{i2}-\mathbf{R}_{j2})}+U\langle{\hat{n}_{k2\sigma}}\rangle+3V\langle{\hat{n}_{k1}}\rangle+D+\Delta,
    \label{eq:epsilon_22}
\end{equation}
while the antidiagonal elements include the single particle contribution ($\sim t^{l\bar{l}}_{ij}$) and the interorbital spin-flip hopping terms (Fock renormalization) originating from the Coulomb interaction and is given by
\begin{equation}
    \varepsilon_{\mathbf{k}\sigma\bar{\sigma}}^{l\bar{l}}=\sum_{\langle i(j) \rangle}(t^{l\bar{l}}_{ij\sigma\bar{\sigma}}-
    V\mathcal{P}^{\bar{l}l}_{j\sigma i\bar\sigma})e^{i\mathbf{k}(\mathbf{R}_{il}-\mathbf{R}_{j\bar l})},
    \label{eq:epsilon_12}
\end{equation}
where $\mathbf{R}_{il}$ vector corresponds to lattice site $i$ on the sublattice $l$ and for $l=1$ ($l=2$) we have $\bar{l}=2$ ($\bar{l}=1$). It should be noted that within the mean-field picture the $V$-term can lead to two effects. One of them corresponds to the onsite energy contribution, which effectively shifts the bands and as a result modifies the energy gap between them [cf. Eqs. (\ref{eq:epsilon_11}) and (\ref{eq:epsilon_22})]. This acts like an additional displacement field tuned by $V$. In addition, a contribution to the NN hopping terms appears with an effective amplitude equal to $V\mathcal{P}^{ll'}_{i\sigma j\bar{\sigma}}$ [cf. Eq. (\ref{eq:epsilon_12})]. Due to the specifics of the considered band structure, the spin-conserving HF contribution of the density-density Coulomb interaction term cannot be operative and therefore is not taken in account. 

In our previous work \cite{chern_1} we considered the case of one hole per moir\'{e} unit cell which leads to a half-filled upper band. In such situation the onsite Coulomb term leads to antiferromagnetism which comes with the cost of losing spin rotational invariance and splitting of the upper band into two sub-bands. Therefore, an antiferromagnetic charge transfer insulating state is realized which after band inversion  generates QAHI state. In contrast, the present study corresponds to two holes per moir\'{e} unit cell meaning that a fully filled lower band and fully empty upper band appears. Now the spin degrees of freedom cannot easily establish a magnetically ordered arrangement with neighboring sites. Therefore, we do not consider any spin-ordering here.

A key aspect of the considered system is that the spin-up valley in MoTe$_2$ is located at the opposite $K$-point of the monolayer Brillouin zone with respect to the spin-up valley in WSe$_2$ as in Ref.\cite{louk}. Therefore, after the band inversion induced by the displacement field, a topological gap can only be opened by interlayer terms, which couple distinct valleys. Due to spin-valley locking these terms are of spin-flip character. For the sake of generality, our model contains such terms already in the single-particle part of the Hamiltonian. However, as already mentioned according to lowest order continuum modeling, the spin-flip hopping amplitudes would be vanishingly small\;\cite{chern_4,Qimao_2024,exp_top_3}. Therefore, here we propose that long-range interactions alone can mediate an effective interlayer hopping mechanism coupling orbitals from different layers and eventually giving rise to topologically nontrivial bands even when the hybridization term is absent in the single-particle Hamiltonian. As can be seen from Eq. (\ref{eq:epsilon_12}) such mechanism can already appear at the mean-field level. However, for the case of $t_{\perp}=0$, this would require spontaneous symmetry breaking leading to a situation in which ${{\mathcal{P}}^{ll'}_{i\sigma j\bar{\sigma}}}\neq 0$ (for $l\neq l'$) results due to interactions. In such a case the Coulomb repulsion together with the specific type of spin-valley locking would lead to an emergent topological state. The analysis of such a scenario is the starting point of the current work.

\section{\label{sec:level3}Results}
Our analysis focuses on the extended Kane-Mele-Hubbard model applied to the AB-stacked MoTe$_2$/WSe$_2$ heterobilayer. We employ a self-consistent HF mean-field approach at the integer filling of two holes per moir\'{e} unit cell \( v = 2 \), for which experiments have revealed a continuous topological phase transition from a moir\'{e} band insulator to a Quantum Valley Hall Insulator. Within our two-band picture the experimental situation of \( v = 2 \) corresponds to the lower band being completely filled and the upper band being empty. In the following, we use the electron language in which the number of electrons per moir\'{e} unit cell is $n=4-v$ , meaning that $v=2$ corresponds to $n=2$.

\subsection{QVHI state formation}\textbf{}

First, we analyze the possibility of creating a topological state purely via Coulomb-assisted spin-flip hopping when $t_{\perp}=0$ and $\mathbf{B}=0$. The topological gap can be opened by antidiagonal hybridization terms in the Hamiltonian matrix (\ref{eq:matrix_hamiltonian}) when a band inversion appears. As seen from Eq. (\ref{eq:epsilon_12}) for $t_{\perp}=0$, this requires $|\mathcal{P}^{ll'}_{i\sigma j\bar{\sigma}}|\neq 0$ for $l\neq l'$. In Figs. \ref{3_bands} (a) and (b) we show that in a certain range of model parameters, such a QVHI state can become stable. In the obtained solution the following rule takes place: $|\mathcal{P}^{ll'}_{i\sigma j\bar{\sigma}}|=\mathcal{P}_{\perp}\in\mathbb{R}$ for $l\neq l'$ for all six possible NN pairs. Therefore, the mixing term induced by the Coulomb interactions does not fully resemble the one that appeared in the original single-particle Hamiltonian [cf. Eq. (\ref{hopping_amp_interlayer})]. Nevertheless, it still leads to non-trivial topological features; hence, the non-zero values of $\mathcal{P}_{\perp}$ lead to opening of a topological gap, and the Chern number being $C=1$ and $C=-1$ corresponding to the two valleys $K$ and $K'$, respectively. Due to purely real NN hopping amplitudes in all directions, the resulting topological gap corresponds to the $A_1$ irreducible representation ($s$-$wave$ symmetry). For completeness, we provide the calculated Berry curvature in the $\mathbf{k}$-space for the obtained QVHI state in Appendix B.

One important remark is in place here. Namely, in the considered model, the two opposite values of the Chern number do not correspond to spin-up and spin-down subbands, but to $K$ and $K'$ valleys. In this sense, we call the obtained state a Quantum Valley Hall state instead of Quantum Spin Hall state. In other words, the band gap seen in Fig. \ref{3_bands}(b) is opened due to a nearest-neighbor spin-flip terms between two opposite-spin bands. Hence, the edge states cannot be the eigenstates of the $\hat{S_z}$ operator. However, the two counter propagating states on a given edge still form Kramers pairs, which are symmetry protected. Therefore, the obtained global topological invariant is $Z_2=1$. This constitutes a fundamental difference in comparison with the Quantum Valley Hall Insulator created by adding a staggered onsite potential to a simple graphene-like model\;\cite{Qian_2009,Lee_2020}, where opposite Berry curvature appears around the $K$ and $K'$ points; however, the resulting edge states are not Kramers pairs and globally one gets $Z_2=0$.

Also, by carrying out a unitary transformation, which flips the spins on one of the sublattices, one can transform the model considered here to a model in which the two counter propagating states on each edge correspond to opposite spins which is a cannonical situation known from the Kane-Mele model without the Rashba term\cite{km2}. Since the unitary transformation cannot change the topological invariant, both models have to be characterized by $Z_2=1$.

As one can see in Fig. \ref{3_bands} (b), to induce $\mathcal{P}_{\perp}\neq 0$ it is necessary to have significant Coulomb repulsion since the effective spin-flip term is proportional to $V$. However, as can be deduced from Eqs. (\ref{eq:epsilon_11}) and (\ref{eq:epsilon_22}), too large values of $V$ increase the energy gap between the bands originating from the two layers. This can make the interlayer mixing term not operative, leading to the suppression of $\mathcal{P}_{\perp}$ and consequently to the destruction of the topological state. Additionally, since the displacement field also tunes the gap between the bands, it has to be in proper range so that the interlayer mixing is possible as visible in Fig \ref{3_bands}(a). Initially, by increasing $D$ one moves the bands closer to each other which works in favor of generating nonzero values of $\mathcal{P}_{\perp}$ . However, too large values of $D$ lead to the WSe$_2$ band being placed above the band originating from the MoTe$_2$ layer. In such a regime, further increase of $D$ enhances the band gap which has a negative influence on the mixing and suppresses $\mathcal{P}_{\perp}$.

To better understand the sequence of phases shown in Fig. \ref{3_bands} we present the band structure for three representative values of the gate voltage values (\(D=45\)meV, \(D=55\) meV, \(D=70\)meV) each corresponding to Coulomb interaction \(V\) = \(19\) meV and \(U\) = \(94.7\) meV. With increasing $D$ one encounters a following phases: (i) moir\'{e} band insulator with the MoTe$_2$ and WSe$_2$ bands being separated by a significant band gap with $\mathcal{P}_{\perp}=0$; (ii) QVHI state where the band inversion appears together with $\mathcal{P}_{\perp}\neq 0$ leading non-trivial topology; (iii) metallic state in which the spin-flip terms are again suppressed to zero. Note, that the continuous phase transition from band insulator to QVHI obtained by us here is consistent with the key experimental results shown in Refs. \cite{exp_1, b_field}.

It should be noted, that apart from the gap opening at the transition to the topological state, also a charge transfer appears from the $l=2$ orbital to $l=1$ orbital. This generates an energy reduction related with the onsite $\sim U$ term and an energy enhancement related to the intersite Coulomb repulsion $\sim V$ in comparison with the trivial $\mathcal{P}_{\perp}=0$ state. Additionally, there is a slight single particle energy reduction. In total there is an energy decrease which works in favor of the stabilization of the topological state for the specified parameter regime shown in Fig. 2(a) and (b). More details related with the relative contributions to the energy reduction at the transition to the topological state, are provided in Appendix B.

\subsection{Symmetry of the topological state}
\begin{figure}[t]
\centering
\includegraphics[width=0.9\linewidth, height=7.2cm]{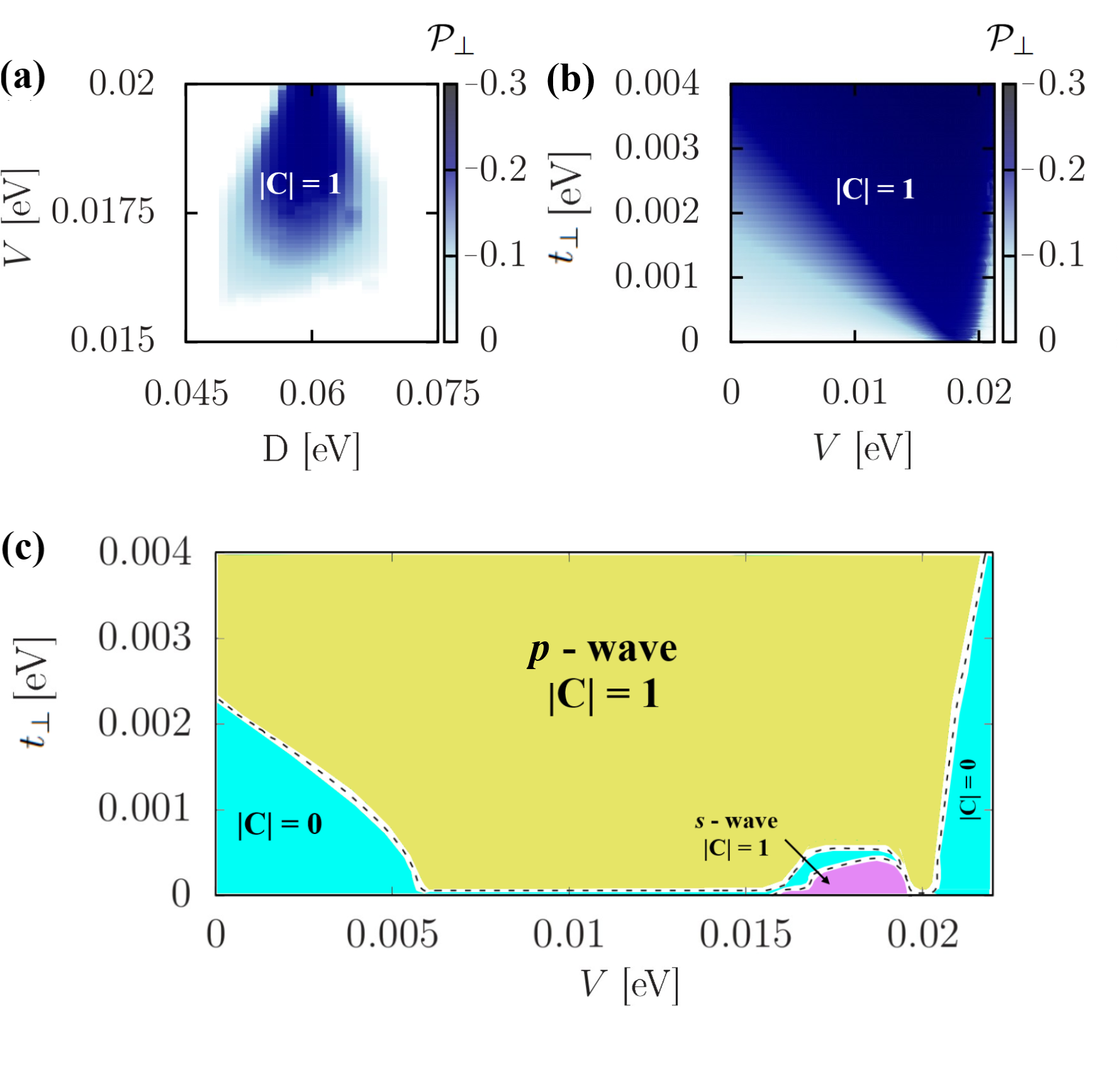}
\caption{(a) The interlayer tunneling expectation value $\mathcal{P}_{\perp}$ as function of the displacement field ($D$) and intersite Coulomb interaction ($V$) for $t_{\perp}=0$. (b)  The interlayer tunneling expectation value $\mathcal{P}_{\perp}$ as function of $V$ and $t_{\perp}$ for selected displacement field $D$ = 55 meV and $\phi_{\perp}=0$. (c) The phase diagram of the model at the ($t_{\perp}, V$)-plane for $D$ = 55 meV and $\phi_{\perp}=2\pi/3$. The two topological states seen in (c) correspond to a \(p\pm ip\) and $s$-$wave$ symmetries of the gap.
}
\label{2_blue}
\end{figure}

In Fig. \ref{2_blue}(a), we provide the calculated interlayer spin-flip expectation value $\mathcal{P}_{\perp}$ which leads to topological gap opening in certain region of the $(D,V)$-plane. This region corresponds to QVHI state with $C=\pm 1$. As one can see, $V_{\mathrm{opt}}\approx18-19$\;meV corresponds to relatively large topological gap and the appearance of the QVHI state, which remains stable within a displacement field window between approximately \(50\) meV and \(70\) meV. It should be noted that the determined optimal value of $V$ is comparable but smaller than the width of the single band of the model ($W\approx25-35$\;meV). Additionally, in Fig. \ref{2_blue} (b) and (c) we consider a situation in which the interaction induced NN hopping is supplemented with a contribution resulting from the single-particle part of our Hamiltonian. For purely real single-particle hopping amplitude ($t_{\perp}\neq 0$ and $\phi_{\perp}=0$) the effect is rather trivial. Namely, increasing $t_{\perp}$ stabilizes the topological phase, making it more robust in the diagram. In such case, even for $V=0$ one can induce the $|C|\neq0$ situation.

An interesting situation takes place when the contribution to the NN hopping originating from the single-particle terms is characterized by a complex amplitude with $\phi_{\perp}=2\pi/3$ [cf. Eq. (\ref{hopping_amp_interlayer})]. Such particular phase appearing in the single particle part of the Hamiltonian has been considered in Ref. \cite{louk}. As one can see from Fig. \ref{2_blue} (c), in such a case a competition between two topological states appears. The first is characterized by an $s$-$wave$ symmetry of the topological gap, while in the second a $p\pm ip$ symmetry is realized. The $s$-$wave$ state is promoted by the interaction-based mechanism analyzed earlier and appears in a relatively narrow window of $V$ and very small $t_{\perp}$ amplitude. As $t_{\perp}$ increases, the tendency towards the $p\pm ip$ symmetry arises and at some point the NN tunneling expectation values ($\mathcal{P}^{ll'}_{i\sigma j\bar{\sigma}}$ for $l\neq l'$) start to reproduce the complex phase pattern originating from the single particle term. However, there is a narrow transition region in which $|C|=0$. 


\begin{figure}[t]
\centering
\includegraphics[width=1\linewidth, height=8.5cm]{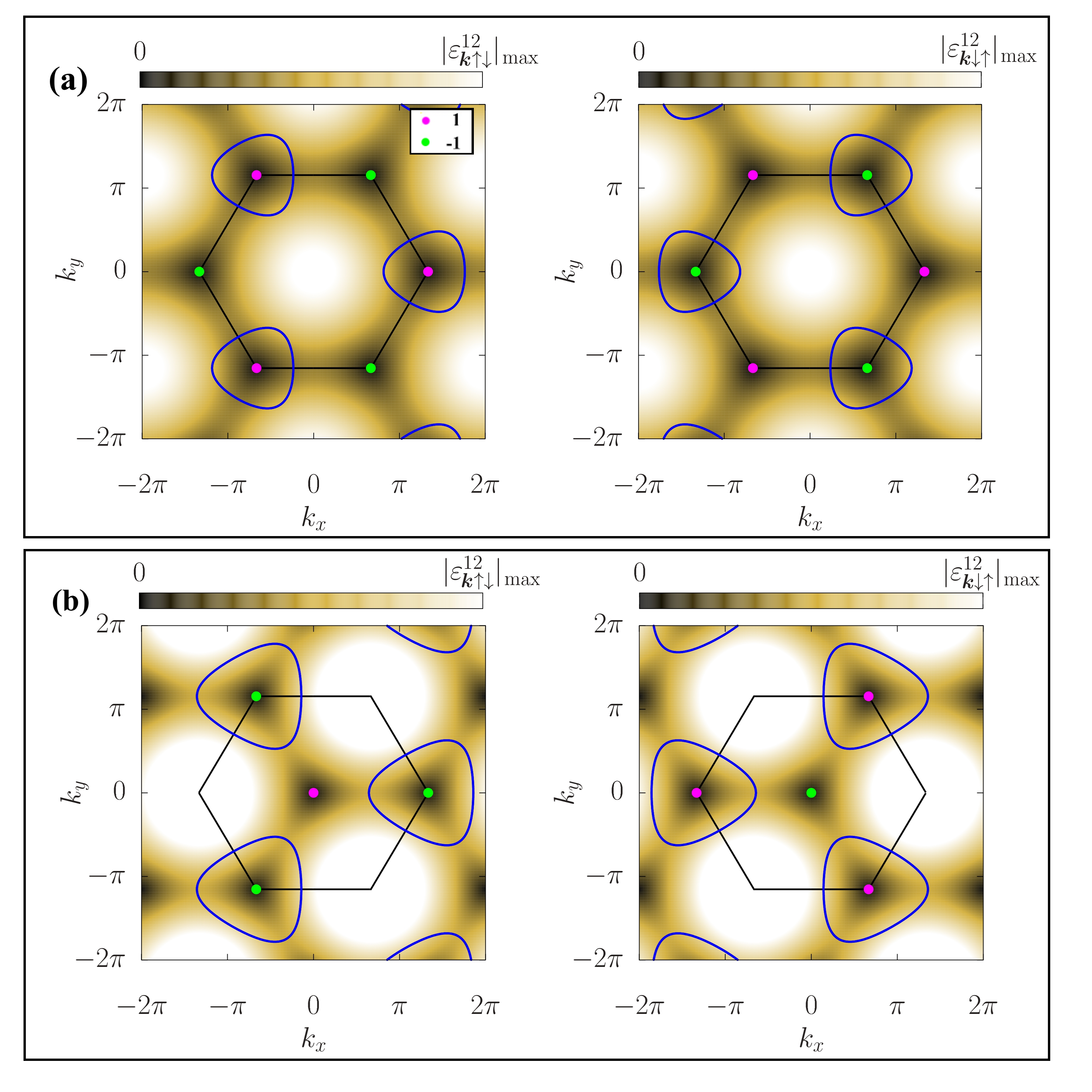}
\caption{Absolute value of the topological gap ($|\varepsilon_{\mathbf{k}\sigma\bar\sigma}^{12}|$) as a function of momentum for pure $s$-$wave$ (a) and $p\pm ip$-$wave$ (b) symmetries. The magenta and green dots mark the positions of the Chern charges of the value $1$ and $-1$, respectively. (a) and (b) correspond to $t_{\perp}=0$ and $t_{\perp}=3\;$meV with $\phi_{\perp}=2\pi/3$. The remaining model parameters are $V=19$\;meV, $D=55$\;meV and $U=94.7$\;meV.
 }
\label{2_fermi}
\end{figure}

\begin{figure*}[t]
\centering
\includegraphics[width=0.9\linewidth, height=7.25cm]{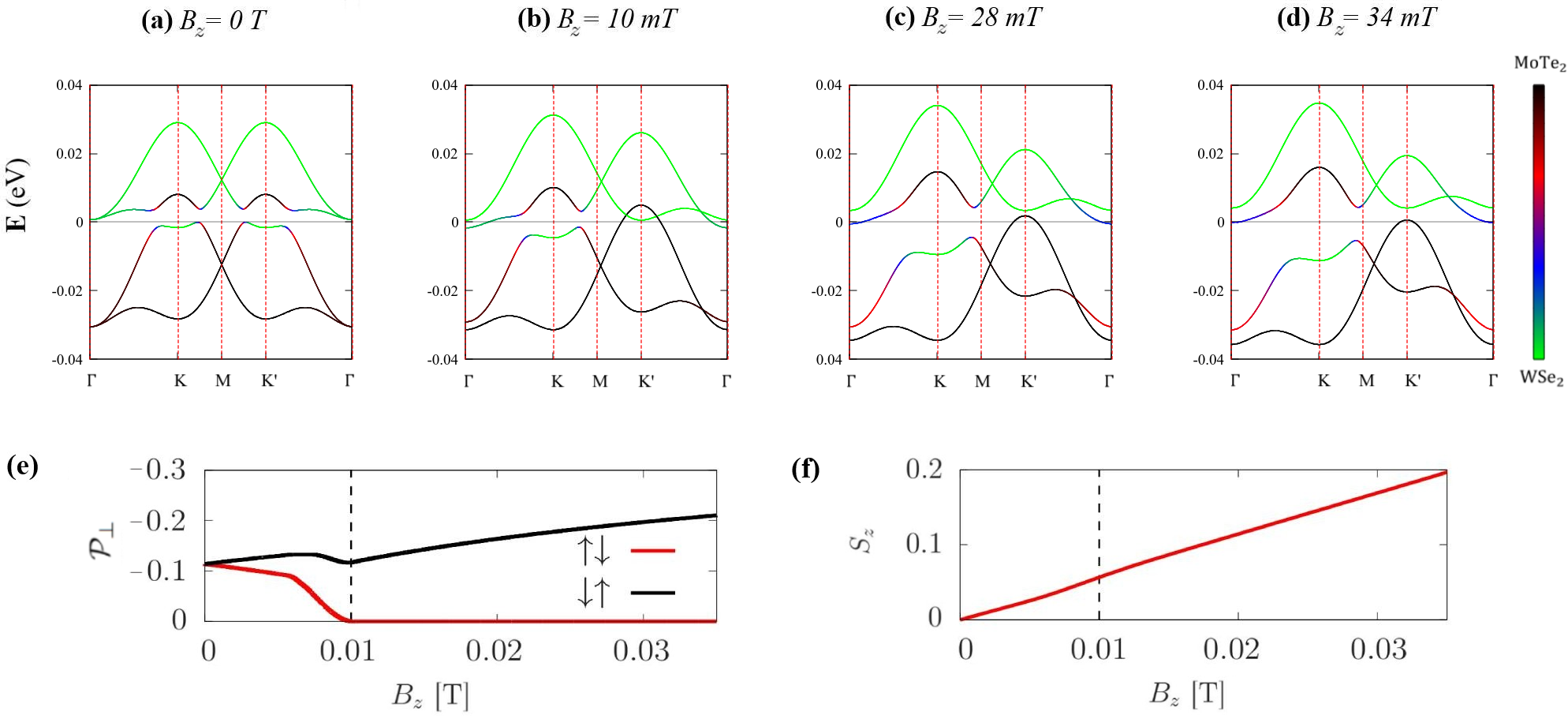}
\caption{(a-d) Band structure along high symmetry directions for selected values of the Zeeman field provided in the figures and for \(D\) = 73 meV, \(U\) = 129 meV, \(V\) = 24 meV. In (e) and (f) we show the $B_z$-dependence of interlayer mixing expectation values $|\mathcal{P}^{12}_{i\uparrow j\downarrow}|\equiv\mathcal{P}_{\perp}^{\uparrow\downarrow}$ and $|\mathcal{P}^{12}_{i\downarrow j\uparrow}|\equiv\mathcal{P}_{\perp}^{\downarrow\uparrow}$ as well as the spin magnetization \(S_z\).
}
\label{qahi_QVHI}
\end{figure*}

To gain more insight when it comes to the differences between the two obtained topological states, in Fig. \ref{2_fermi} we provide the color maps showing the absolute value of the $\mathbf{k}$-dependent topological gaps, $|\varepsilon_{\mathbf{k}\sigma\bar\sigma}^{12}|$. Note, that the gap opening is caused by the anti-diagonal terms appearing in the matrix form of our Hamiltonian [cf. Eq. (\ref{eq:matrix_hamiltonian})]. One of the visible differences is that in the $s$-$wave$ scenario the absolute value of the gap is $C_6$ symmetric and does not depend on the spin configuration, while in the $p$-$wave$ case a two $2\pi/3$ rotated patterns are created corresponding to the two spin configurations, which are $C_3$ symmetric each. In both situations, nodal points appear at the $K$ and $K'$ points  for which $|\varepsilon^{12}_{\mathbf{k}\sigma\bar{\sigma}}|=0$. However, for the $p\pm ip$ symmetry, there is an additional nodal point at $\Gamma$. Note that each nodal point introduces a topological charge and the Chern number can be calculated by using the following formula
\begin{equation}
    C=\sum_{i\in \Omega_{F}}(-1)^{w}C_{i},
\end{equation}
where \(C_i\) is the topological charge of a given nodal point which can be determined by calculating the winding of the complex phase of $\varepsilon^{12}_{\mathbf{k}\sigma\bar\sigma}$ around it with \(w=0\;(1)\) for the clockwise (anticlockwise) direction. It is important to note that  index \(i\) in the summation above runs over the nodal points contained inside the closed curve determined by the condition $\varepsilon^{11}_{\mathbf{k}\sigma}=\varepsilon^{22}_{\mathbf{k}\bar{\sigma}}$ (cf. Appendix C). The values of the nodal points and the $C_F$ curves are provided in Fig. \ref{2_fermi}. It can be clearly seen that both scenarios lead to the Chern number $|C|=1$ for each valley. Due to spin-valley locking the two valleys are characterized by opposite Chern numbers which leads to QVHI state. Note that the topological charge of each $K$ and $K'$ point has to be divided by 3 since it is shared by three Brillouin zones. The values of the Chern numbers has been additionally calculated by using the Brillouin zone triangulation method which gives the same results as the procedure described above.

\subsection{Influence of Magnetic Field}

Following the recent experimental report \cite{b_field} we try to replicate the proposed topological phase transition induced by the presence of external magnetic field within the Coulomb-assisted hopping scenario presented by us in the previous Section (for $t_{\perp}=0$). Within the considered model, the Zeeman term can tune the topological gaps appearing at the two valleys of the mini Brillouin zone. However, with increasing $B_z$, the gap in one valley is enhanced while the other is suppressed. 

In this part of our analysis we slightly change the model parameters $U$ and $V$ so as the transition to the QAHI state is clearly visible with increasing magnetic field. Optimum values of $U$ and $D$ are chosen so that the band inversion is comparatively lower than the one in Fig. \ref{3_bands}(d) and the topological gap is reduced which is a perfect recipe for the topological transition shown here. Also, for larger $V$ taken here, the orbital splitting induced by the interactions suppresses the hybridization between the two bands which is needed to terminate mixing in one of the valleys to enter QAHI phase. We start with the following model parameters \(U\) = 129 meV, \(V\) = 24 meV, and \(D\) = 73 meV which lead to a topological gap opened as shown in Fig. \ref{qahi_QVHI}(a) for $B_z=0$. Similarly as before in such a situation we have $\mathcal{P}^{ll'}_{i\sigma j\bar\sigma}\equiv\mathcal{P}_{\perp}\neq 0$. However, under the application of small nonzero \(B_z\) the band inversion is reduced in the $K'$ valley, and consequently $|\mathcal{P}^{12}_{i\uparrow j\downarrow}|\equiv\mathcal{P}_{\perp}^{\uparrow\downarrow}$ is significantly suppressed since it corresponds to band mixing in the same valley  [cf. Fig. \ref{qahi_QVHI}(d)]. At the same time, the opposite situation takes place in the $K$ valley where the band inversion is enhanced together with the value of $|\mathcal{P}^{12}_{i\downarrow j\uparrow}|\equiv\mathcal{P}_{\perp}^{\downarrow\uparrow}$. For the critical value of $B_z=10$ mT some small band inversion at $K'$ still appears but the corresponding Coulomb assistant hopping is already zero. This leads to a loss of topological properties of one-half of the considered Kane-Mele-Hubbard model. Nevertheless, the $K$-valley gap is still topological. In such a state, a combination of topological insulating and normal metallic behavior appears. Further enhancement of the Zeeman term leads to opening of a trivial band gap at $K'$ and enhancement of the topological gap at $K$ meaning that a QAHI state should be observed [cf. Fig. \ref{qahi_QVHI}(d)]. As a consequence under the influence of Zeeman effect we have a transition from QVHI state to QAHI state under fixed bias and interaction strengths (both on-site and inter-site).


\section{Conclusions}
In summary, we carry out a detailed theoretical investigation of AB-stacked MoTe$_2$/WSe$_2$ heterostructure at $\nu=2$ and propose that the non-trivial topology in this structure can be driven by many body interactions. We show that the Coulomb interaction-assisted interlayer hopping can lead to the opening of an $s$-$wave$ topological gap leading to a QVHI state even in the absence of the single-particle band mixing term. In order to create favorable conditions for such a spontaneously formed topological state, the NN Coulomb interactions must be relatively strong, with the optimal value being $V_{\mathrm{opt}}=18$-$19\;$meV. In such a situation, by increasing the displacement field one encounters the following sequence of phases: (i) moir\'{e} band insulator; (ii) QVHI state (iii) metallic state. The continuous phase transition from band insulator to QVHI obtained here is consistent with the key experimental results shown in Refs. \cite{exp_1, b_field}.



It should be noted that the interlayer hopping amplitudes ($t_{\perp}$) originating from the single particle part of our Hamiltonian for the AB-stacked MoTe$_2$/WTe$_2$ are anticipated to be very weak or vanishingly small\cite{chern_4,Qimao_2024,exp_top_3,exciton_2}. As mentioned, in the first part of our analysis we focus on the case of $t_{\perp}=0$ and show that a spontaneously formed  $s$-$wave$ topological state can be formed purely due to the NN Coulomb interactions. For the sake of completeness, in the second part of our study we also consider a situation in which the band mixing term originates both from the interaction induced mechanism and due to the single-particle physics. As seen in Fig. \ref{2_blue}(c) for the regime of small interlayer hopping, $t_{\perp}\lesssim1\;$meV,  which still can be considered as realistic both $s$-$wave$ and and $p\pm ip$-wave symmetries of the topological gap are possible depending on the value of the displacement field. The main difference between the two states is that in the $p\pm ip$-wave case the $\mathbf{k}$-dependent topological gap has an additional Chern charge residing at the $\Gamma$ point, which is absent for the $s$-$wave$ state. However, since in both situations the Chern charges of the $K$ and $K'$ points determine the topological features, the two situations lead to QVHI with $C=\pm 1$.

Finally, we show that the topological phase transition induced by the presence of external magnetic field analyzed in experiments\cite{b_field} can be reproduced within the Coulomb-assisted hopping scenario considered by us. In such a case, the application magnetic field produces trivial gap at one valley with enhancement of topological gap at the other one thereby favoring the realization of a QAHI phase.

The data behind the figures are available in the open repository\;\cite{zegrodnik_michal_2025_zenodo}.


\begin{acknowledgments}
We acknowledge discussions with Louk Rademaker and Julian Czarnecki. This  research was partly supported by National Science Centre, Poland (NCN) according to Decision No. 2021/42/E/ST3/00128. 


\end{acknowledgments}

\appendix

\section{Flavor Symmetry}
It should be noted that before any emergence of hybridization strong Ising spin–orbit coupling locks spin to valley ($\tau$) in each layer (\(l\)) making the two degrees of freedom effectively interchangeable. The flavor space of the model is thus created by the (\(l\) \(\otimes\) \(\tau\)) subspace which is appropriate for the low energy physics. Hence we get a 4 dimensional flavor space spanned by \{$(1K),(1K'),(2K),(2K')$\}. Due to lack of inter-flavor coupling, each flavor conserves it's own particle number which is why the flavor-space Hamiltonian is invariant under \(U(1)_{1K}\) $\times$ \(U(1)_{2K}\) $\times$ \(U(1)_{1K'}\) $\times$ \(U(1)_{2K'}\) symmetry. 

If \(|t_{\perp}|\) $\neq$ \(0\) onset of spin-conserving interlayer mixing is still able to conserve separate particle numbers in each valley but fails to do the same for layer indices lowering the symmetry to \(U(1)_{K\sigma}\) $\times$ \(U(1)_{K'\bar\sigma}\). The special case in which the single particle physics of this heterobilayer is worked out has only finite spin flip hybridization ($\mathcal{P}_{\perp}$) thus destroying the spin-valley locked feature but with the same symmetry of \(U(1)_{K}\) $\times$ \(U(1)_{K'}\) without the spin indices.

\begin{table}[b]
\caption{
 The table presents the total energy difference per unit cell between the topological state with \( P_\perp \neq 0 \) and the trivial state with \( P_\perp = 0 \). The total energy difference ($\Delta E_{tot}$) reflects the combined contribution from all energy terms provided in Eqs. (B2) and (B4) for the situation discussed in Fig. 2(d) (\(D\) = $55$ meV, \(V\) = 19 meV and \(U\) = 94.7 meV), for which the \( P_\perp \neq 0 \) state is stable. Negative values mean that the given tern works in favor of the stabilization of the topological state with $\mathcal{P}_{\perp}\neq 0$, while the positive values correspond to an energy penalty.}
\vspace{0.45cm}
\label{tab:energy}
\centering
\setlength{\tabcolsep}{10pt}
\begin{tabular}{ccc} 
 \hline\hline
 & $ E (\mathcal{P}_{\perp}\neq  0 ) - E(\mathcal{P}_{\perp} =  0)\;$[meV] \\ 
 \hline  
 \\[-6pt]
 $\Delta E_{t}$ & -0.48\\ 
 $\Delta E_{0}$ & 0.66  \\ 
 $\Delta E_{U}$& -8.04 \\ 
 $\Delta E_{V}$& 9.87  \\ 
 $\Delta E_{V}'$& -2.45  \\ 
  $\Delta E_{tot}$ & -0.44 \\[0.2cm]
 \hline\hline
\end{tabular}
\end{table}

\section{Energy reduction at the transition to the topological state}
To interpret the spontaneous formation of the topological state with $\mathcal{P}_{\perp}\neq0$ shown in Sec. IIIA for $t_{\perp}=0$, we write down the mean field expression for the subsequent energy terms of our Hamiltonian. The single particle energy per unit cell can be cast in the following form
\begin{equation}
\langle \mathcal{H}_t\rangle/N = E_t + E_0,\\
\end{equation}
where the hopping and onsite contributions are
\begin{equation}
\begin{split}
E_t &= \frac{1}{N}\sum_{\langle\langle ij\rangle\rangle l\sigma} t^{ll}_{ij\sigma}\;\langle\hat{c}_{i l \sigma}^{\dagger}\; \hat{c}_{j l \sigma}^{}\rangle ,\\
E_0 &= (D+\Delta)(n+\delta n),\\
\end{split}
\end{equation}
and $N$ is the number of unit cells in the system, while $\langle\hat{c}^{\dagger}_{il\sigma}\hat{c}_{il\sigma}\rangle\equiv n_l/2$ for $l=1,2$ and $n_1=n-\delta n$, $n_2=n+\delta n$ since we are considering a non-magnetic state and the onsite energy offset ($\Delta$) and displacement field ($D$) contribution leads to a charge imbalance $\delta n>0$ between the two orbitals of the system.
The interaction energy per unit cell has the following form
\begin{equation}
    \langle \mathcal{H}_{UV}\rangle/N=E_U+E_V+E_V',
\end{equation}
where
\begin{equation}
\begin{split}
E_U &=\;\frac{U}{2}(n^2+\delta n^2),\\
E_V &=3\;V(n^2-\delta n^2), \\
E'_V &=-6\;V|\mathcal{P}_{\perp}|^2.\\
\end{split}
\label{eq:E_UV_contributions}
\end{equation}
For the specific range of model parameters $V$ and $D$ shown in Figs. 2 and 3, the topological state with $\mathcal{P}_{\perp}\neq0$ becomes stable meaning that its total energy is smaller than the total energy of the trivial solution with $\mathcal{P}_{\perp}=0$. 
In order to intuitively understand such effect, one should note that a band inversion appears in the considered case; therefore, as soon as $\mathcal{P}_{\perp}$ becomes non-zero an effective interband mixing term opens a gap and leads to a charge transfer from $l=2$ orbital to $l=1$ orbital. That makes $\delta n$ decrease at the transition to the $\mathcal{P}_{\perp}\neq0$ state. From Eqs. (\ref{eq:E_UV_contributions}) one can see that such charge transfer decreases the $E_U$ energy but increases the $E_V$ energy. At the same time $E'_V$ has to decrease at the transition since $\mathcal{P}_{\perp}$ becomes non-zero. Therefore, $E_U$ and $E'_V$ terms work in favor of the newly created topological state while $E_V$ generates an energy penalty.   
When it comes to the single particle terms, the hopping term ($\sim \mathcal{P}^{(l)}_{\parallel}$) work slightly in favor of the $\mathcal{P}_{\perp}\neq 0$ state while the onsite single particle term generates energy penalty since $D+\Delta<0$.

In Table I we provide the value of the total energy reduction at the transition to the topological state, $\Delta E=E (\mathcal{P}_{\perp}\neq  0 ) - E(\mathcal{P}_{\perp} =  0)$, together with the particular contributions corresponding to Eqs. (B2) and (B4) for selected set of model parameters. To sum up, the most significant contribution when it comes to stabilization of the topological state originates from the onsite repulsion term ($\sim U$) after the mixing term ($\sim V$) initiates charge transfer between the two orbitals.

\section{Chern number as a winding number}
Here we provide details related with the characterization of the topological features of the considered model based on the winding number. We start with rewriting the matrix form of our Hamiltonian in a compact form
\begin{equation}
    \hat{H} = \sum_{\mathbf{k}\sigma}\hat{\mathbf{f}}^{\dagger}_{\mathbf{k}\sigma}\mathbf{H}_{\mathbf{k}\sigma}\hat{\mathbf{f}}_{\mathbf{k}\sigma}
    \label{eq:Hamiltonian_Appendix}
\end{equation}
where $\hat{\mathbf{f}}^{\dagger}_{\mathbf{k}\sigma}=(\hat{c}_{\mathbf{k}1\sigma}^\dagger,\;\hat{c}_{\mathbf{k}2\bar{\sigma}}^\dagger)$ and
\begin{equation}
\mathbf{H}_{\mathbf{k}\sigma} =
\begin{pmatrix}
\varepsilon^{11}_{\mathbf{k}\sigma} & \varepsilon^{12}_{\mathbf{k}\sigma\bar\sigma} \\
\varepsilon^{21}_{\mathbf{k}\bar\sigma\sigma} & \varepsilon^{22}_{\mathbf{k}\bar\sigma}
\end{pmatrix}.
\label{eq:matrix_hamiltonian_appendix}
\end{equation}
In the above Hamiltonian we have omitted the trivial non-operator terms since they are not important in the context considered here. The eigenvalues of the Hamiltonian matrix are the following
\begin{equation}
\lambda^{(\pm)}_{\mathbf{k}\sigma} =
\frac{\varepsilon^{11}_{\mathbf{k}\sigma} + \varepsilon^{22}_{\mathbf{k}\bar\sigma}}{2}
\pm
\delta_{\mathbf{k}\sigma},
\end{equation}
where
\begin{equation}
\delta_{\mathbf{k}\sigma} = \sqrt{ \zeta_{\mathbf{k}\sigma}^2 + |\varepsilon^{12}_{\mathbf{k}\sigma\bar\sigma}|^2 }, \quad
\zeta_{\mathbf{k}\sigma} = \frac{\varepsilon^{11}_{\mathbf{k}\sigma} - \varepsilon^{22}_{\mathbf{k}\bar\sigma}}{2}.
\label{eq:dzeta_and_delta}
\end{equation}
The corresponding eigenvectors can be expressed in the form
\begin{equation}
\begin{split}
\mathrm{for}\quad \lambda^{(+)}_{\mathbf{k}\sigma}:&\quad\ket{u^{(+)}_{\mathbf{k}\sigma}} =
\begin{pmatrix}
e^{i\phi_{\mathbf{k}\sigma}} \cos \frac{\theta_{\mathbf{k}\sigma}}{2} \\[4pt]
\sin \frac{\theta_{\mathbf{k}\sigma}}{2}
\end{pmatrix},\\
\mathrm{for}\quad \lambda^{(-)}_{\mathbf{k}\sigma}:&\quad \ket{u^{(-)}_{\mathbf{k}\sigma}} =
\begin{pmatrix}
e^{i(\phi_{\mathbf{k}\sigma} - \pi)} \sin \frac{\theta_{\mathbf{k}\sigma}}{2} \\[4pt]
\cos \frac{\theta_{\mathbf{k}\sigma}}{2}
\end{pmatrix},\\
\label{eq:eigenvectors}
\end{split}
\end{equation}
with the parameters $\theta_{\mathbf{k}\sigma}$ and $\phi_{\mathbf{k}\sigma}$ defined as
\begin{equation}
\theta_{\mathbf{k}\sigma} = 2\tan^{-1}\!\left( \frac{\delta_{\mathbf{k}\sigma}-\zeta_{\mathbf{k}\sigma}}{|\varepsilon^{12}_{\mathbf{k}\sigma\bar\sigma}|^2} \right),
\qquad
\phi_{\mathbf{k}\sigma} = \arg(\varepsilon^{12}_{\mathbf{k}\sigma\bar\sigma}),
\label{eq:theta_eq}
\end{equation}
where
\begin{equation}
\theta_{\mathbf{k}\sigma} \in [0, \pi], \qquad \phi_{\mathbf{k}\sigma} \in [0, 2\pi].
\end{equation}

\begin{figure}[b]
\centering
\includegraphics[width=0.80\linewidth, height=4cm]{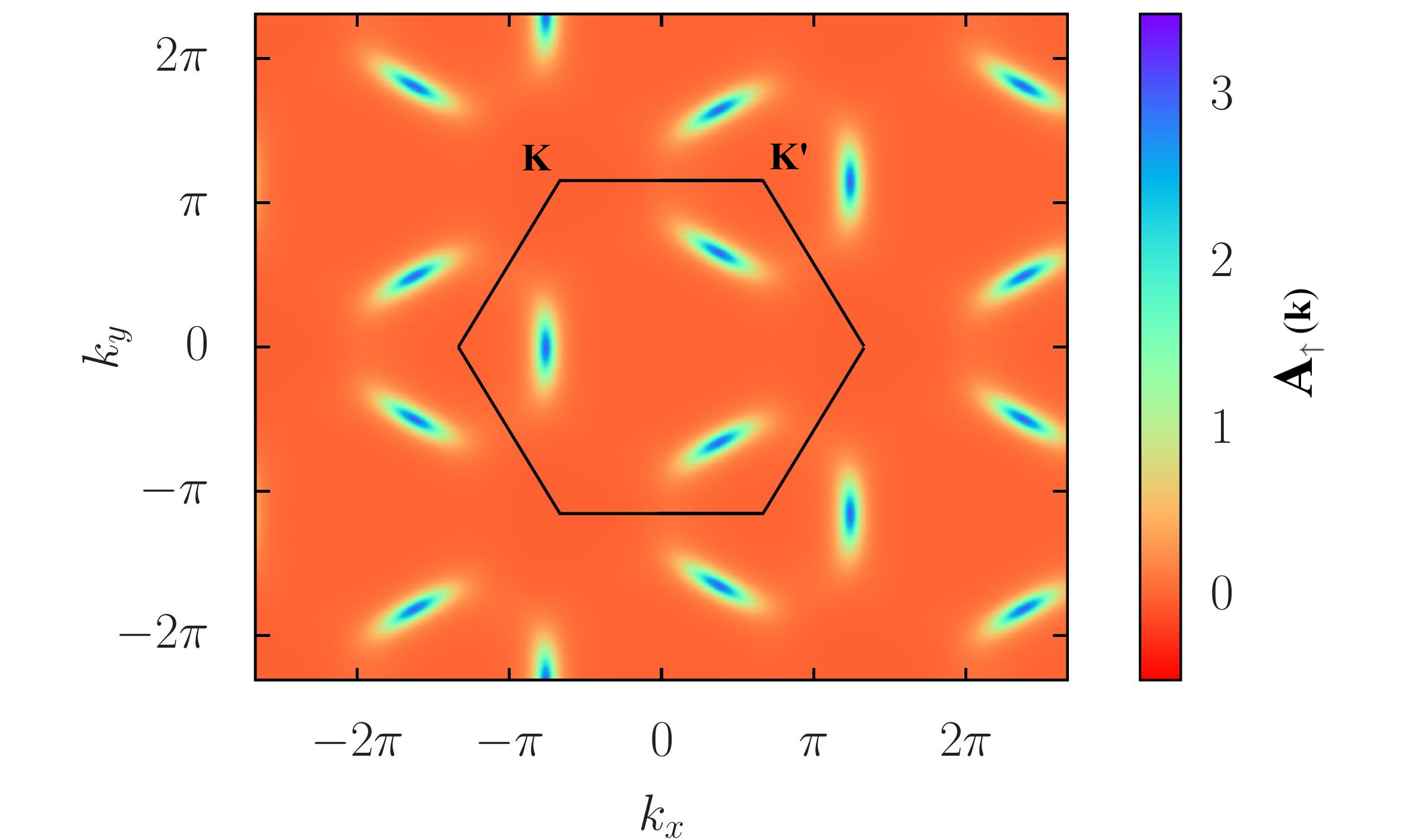}
\caption{Berry curvature in the $\mathbf{k}$-space for $\sigma=\uparrow$
for the case of an \textit{interaction-driven} QVHI state ($t_{\perp}$ = \(0\)) with Chern number $|C|$ = $1$, for the same model parameters as those corresponding to Fig. 2(d) (\(D\) = $55$ meV, \(V\) = 19 meV and \(U\) = 94.7 meV).}
\label{curvature}
\end{figure}
Note, that for a given eigen-state and by using Eqs. (\ref{eq:eigenvectors}) one can map every $\mathbf{k}$ point of the Brillouin zone onto the Bloch sphere with the parameters $\theta_{\mathbf{k}\sigma}$ and $\phi_{\mathbf{k}\sigma}$ interpreted as spherical coordinates and $|2\sigma\rangle$ and $|1\bar{\sigma}\rangle$ states represented by the anti-nodal points. For $n=2$ and the topological gap opened ($\mathcal{P}_{\perp}\neq 0$), the eigenvalue $\lambda^{(-)}_{\mathbf{k}\sigma}$ corresponds to the lower energy band which is fully filled and separated by a band gap from the empty $\lambda^{(+)}_{\mathbf{k}\sigma}$ band. To determine the Chern number we focus on the fully filled $\lambda^{(-)}_{\mathbf{k}\sigma}$ band.

The Chern number is defined as the integral of the Berry curvature over the Brillouin zone. In our case of system, which can be described by the block diagonal Hamiltonian given in Eq. (\ref{eq:Hamiltonian_Appendix}), the Chern number can be determined for each of the blocks representing different spin configurations. Namely,
\begin{equation}
C_{\sigma} = \frac{1}{2\pi} \iint_{\Omega_{\text{BZ}}} \left[ \nabla_{\mathbf{k}} \times \mathbf{A}_{\sigma}(\mathbf{k}) \right] \, d^2k
\label{chern_number}
\end{equation}
where
\begin{equation}
\nabla_{\mathbf{k}} = \left( \frac{\partial}{\partial k_x}, \frac{\partial}{\partial k_y} \right),
\quad
\mathbf{A}_{\sigma}(\mathbf{k}) = i \braket{u^{(-)}_{\mathbf{k}\sigma} | \nabla_{\mathbf{k}} | u^{(-)}_{\mathbf{k}\sigma}}.
\end{equation}
For the sake of completeness, in Fig. \ref{curvature} we show the calculated Berry curvature for the case of interaction induced QVHI state which we discuss in Sec. 3A of the main text. By using Eq. (\ref{chern_number}) and by applying the Brillouin zone triangulation method, one can numerically determine the Chern number. However, we can equivalently calculate the Chern number by using the following equation
\begin{equation}
C_{\sigma} = -\frac{1}{\pi} \oint_{C_F} \mathbf{A}_{\sigma}(\mathbf{k}) \cdot d\mathbf{k}=\frac{1}{2\pi} \oint_{C_F} \nabla_{\mathbf{k}} \phi_{\mathbf{k}\sigma} \cdot d\mathbf{k},
\label{eq:C_winding}
\end{equation}
where the curve $C_F$ over which the integration is carried out corresponds to equal contribution of the $|2\sigma\rangle$ and $|1\bar{\sigma}\rangle$ states in the resultant eigenstate $|u^{(-)}_{\mathbf{k}\sigma}\rangle$. Such a situation appears at the equator of the Bloch sphere, where we are equally distant from the antipodal points, since from Eq. (\ref{eq:eigenvectors}) we have
\begin{equation}
\left| e^{i(\phi_{\mathbf{k}\sigma} - \pi)} \sin \frac{\theta_{\mathbf{k}\sigma}}{2} \right| = \frac{1}{\sqrt{2}},
\qquad
\left| \cos \frac{\theta_{\mathbf{k}\sigma}}{2} \right| = \frac{1}{\sqrt{2}},
\end{equation}
which implies that
\begin{equation}
\theta_{\mathbf{k}\sigma} = \frac{\pi}{2}.
\end{equation}
Equivalently, from the last expression and by using Eqs. (\ref{eq:dzeta_and_delta}) and (\ref{eq:theta_eq}) , one can explicitly determine the condition for the $C_F$ curve in the $\mathbf{k}$-space
\begin{equation}
\epsilon^{11}_{\mathbf{k}\sigma} = \epsilon^{22}_{\mathbf{k}\bar\sigma}
\label{eq:C_F_curve}
\end{equation}
From Eq. (\ref{eq:C_winding}) one can see that the Chern number can be determined by calculating the winding of the complex phase of the hybridization term ($\phi_{\mathbf{k}\sigma}$) around the curve determined by (\ref{eq:C_F_curve}).

\bibliography{refs}

\end{document}